# Phases of physics in J.D. Forbes' *Dissertation Sixth* for the *Encyclopaedia Britannica* (1856)

Isobel Falconer

School of Mathematics and Statistics, University of St Andrews
North Haugh,
St Andrews,
KY16 9SS
UK

Email: ijf3@st-andrews.ac.uk

## Abstract

This paper takes James David Forbes' *Encyclopaedia Britannica* entry, *Dissertation Sixth,* as a lens to examine physics as a cognitive, practical, and social, enterprise. Forbes wrote this survey of eighteenth- and nineteenth-century mathematical and physical sciences, in 1852-6, when British "physics" was at a pivotal point in its history, situated between a discipline identified by its mathematical methods – originating in France - and one identified by its university laboratory institutions. Contemporary encyclopaedias provided a nexus for publishers, the book trade, readers, and men of science, in the formation of physics as a field. Forbes was both a witness, whose account of the progress of physics or natural philosophy can be explored at face value, and an agent, who exploited the opportunity offered by the *Encyclopaedia Britannica* in the mid nineteenth century to enrol the broadly educated public, and scientific collective, illuminating the connection between the definition of physics and its forms of social practice. Forbes used the terms "physics" and "natural philosophy" interchangeably. He portrayed the field as progressed by the natural genius of great men, who curated the discipline within an associational culture that engendered true intellectual spirit. Although this societal mechanism was becoming ineffective, Forbes did not see university institutions as the way forward. Instead, running counter to his friend William Whewell, he advocated inclusion of the mechanical arts (engineering), and a strictly limited role for mathematics. He revealed tensions when the widely accepted discovery-based historiography conflicted with intellectual and moral worth, reflecting a nineteenth-century concern with spirit that cuts across twentieth-century questions about discipline and field.

## Introduction

> The object of this Dissertation has little in common with an attempt
> formally to subdivide human knowledge into compartments.... It is chiefly
> in their practical bearing on one another that they must be considered. If
> one science, like Mathematics, furnish the only sure step towards the
> understanding ... of another, as Astronomy or Optics, a practical link is
> constructed between them.... The intimate and reciprocal connection thus
> subsisting between Mathematics and Physics is to be found in almost an
> equal degree between Pure Physics and the Mechanical Arts, of which we
> take Civil Engineering to represent the department most cognate to that
> of Natural Philosophy....[1]

This characterisation of physics in 1856 by James David Forbes (1809-68),
Professor of Natural Philosophy at Edinburgh, is notable for his refusal to erect
boundaries around physics, his emphasis on relational practice, and his use of
"physics" and "natural philosophy" as interchangeable.

Forbes' *Dissertation sixth: Exhibiting a general view of the progress of
mathematical and physical science, principally from 1775 to 1850,* was written for
the eighth edition of the *Encyclopaedia Britannica.* Chronologically, it fell
between the two main British historiographies of the rise of "physics": right at
the end of the period to which Buchwald and Hong, and Morus, ascribe the
formation of physics through the mathematization of the experimental sciences
of heat, optics, electricity and magnetism.[2] Equally, it just pre-dates the
institutionalisation of physics teaching into university laboratories in Britain,
and the associated social-institutional concerns, that Gooday identifies as key to
discipline formation in the 1860s and '70s.[3]

Forbes' account might, then, be interpreted as an indicator of "physics" at a
pivotal point in its history, situated between a discipline that identified itself by
its methods – originating in France with men such as Joseph Fourier and
Augustin Fresnel, and curated in Britain by Cambridge mathematicians such as
John Herschel, George Peacock, and William Whewell - and one that identified
itself by its institutions. However, Forbes was not an independent observer, he
was an agent in this change, as recognised by scholars of the 1970s and '80s,
such as Wilson, Harman, Olson, and Smith and Crosland, who asserted his
influence on students such as James Clerk Maxwell, Balfour Stewart, Peter
Guthrie Tait, and William John Macquorn Rankine.[4]

Geographically, also, Forbes' situation in the historiography of the changing
constitution of "physics" is ambiguous. Born, educated, and based through most
of his career, in Edinburgh, Forbes seems remote from either account. Although
Morus has countered the Cambridge-centric view of the earlier period by adding
London, and Gooday the Oxbridge dominance of the later by demonstrating the
crucial role of the rapidly expanding industrially driven universities, neither
historiographic account situates the action in Edinburgh.[5] Yet Forbes travelled
frequently and extensively across Britain and the rest of Europe and was an
inveterate networker, exchanging news, books, and instruments, with
Continental workers such as François Arago, Adolphe Quetelet, and Auguste de la
Rive. He was in close contact with Cambridge mathematicians such as Whewell,
Herschel, and George Biddell Airy, becoming one of Morrell and Thackray's

"gentlemen of science" – the powerful coterie responsible for the foundation and subsequent success of the British Association for the Advancement of Science (BAAS).[6] He was both removed from the major powerhouses said to drive the development of institutional physics, but, conversely, deeply embedded among the cultivators of physical science.

Responding to such ambiguities, this paper takes a reflexive approach. It views Forbes both as a witness whose account of the state and progress of physics can be explored at face value, and as an agent whose exploitation of the opportunity offered by the *Encyclopaedia Britannica* in the mid nineteenth century is worth examining for what it tells us about the boundaries of physics and the connection between the definition of physics and its form of social practice.

Yeo associates the shift in emphasis in British encyclopaedias in the early nineteenth century from classification of universal knowledge, to "detailed entries on scientific disciplines by expert contributors", with the emergence of specialised disciplines from natural philosophy.[7] Alongside textbooks and specialised journals, Stichweh identifies encyclopaedia entries, written by specialists and increasingly read by specialists, as the mechanism through which disciplines self-organise. But he acknowledges that this internal process requires supporting external cultural conditions and a critical mass of potential specialists – criteria that were met in the case of German physics.[8] No such conditions applied in Scotland until the mid nineteenth century. Its five universities, with one professor of natural philosophy and one of mathematics in each, were not enough to provide an internal disciplinary physics network divorced from the wider educated elite, until the late nineteenth century. Forbes' own family were typical of the landed gentry, lawyers, and bankers who formed the major part of the local audience for science, especially in Edinburgh where the university was under civic control. The content of mathematico-physical science held little attraction for them and they related better to natural sciences.[9] In reponse, Scottish professors of natural philosophy, whose inadequate "fixed" salaries were supplemented by student fees, retained a curriculum that was generally accessible to all social and professional types, insisting that their subject remain a core part of the "M.A." (first arts degree).[10] This limited the extent to which they could produce specialists in physical science.

Outside the elite, there were three other sources of potential specialists, found in other localities: the medical community, military academies, and the growing artisan group.[11] Although Edinburgh had had a world-leading medical community, this was weakening by the 1820s. In contrast to London, natural philosophy was no longer a part of its medical degree and the plethora of public lectures of the mid eighteenth century had given way to chemistry by around 1800.[12] Nor were there military academies nearby: Scottish mathematicians such as James Ivory and William Wallace had to move south for such employment. So, unlike the French situation described by Simon, or that around London, there was little demand for circulation of medically- or militarily-oriented textbooks to prompt formation of physics as a discipline in Scotland.[13] Perhaps more significant, but under-explored by Scottish historians, was the growing nexus of science- and technology-oriented colleges, improvement societies, and mechanics institutes for artisans.[14] By 1851 there were 55 adult education institutions of some type across Scotland, with over 12,500 members. England had 610, and over 102000 members.[15] The Edinburgh School of Arts awarded a Diploma of Life Membership on successful completion of a three-year course in mathematics, chemistry, and natural philosophy; from 1835 to 1851 they awarded 46.[16] In 1824, three years after the Edinburgh School of Arts was founded, a commentator declared, "A spirit has been awakened, which, if properly directed, may be productive of the most beneficial effects…."[17] The *Encyclopaedia Britannica* provided an opportunity for the Edinburgh-based Forbes to supply "proper direction" to artisans, while simultaneously both shaping the established community of cultivators of physical science and enrolling the educated elite into the values of scientific men and philosophers as he saw them.[18] His *Dissertation* can be read both as a testament of what physics was during the first half of the nineteenth century, and as a manifesto for the second half.

When he wrote *Dissertation Sixth*, Forbes was in his mid 40s, and well established figure in British science. The over-protected youngest son of an influential Scottish banking family, he had decided early on a career in physical science and pursued his goal relentlessly through a combination of dedicated hard work and family influence. In 1833, aged 23, he mounted an intensely political campaign that saw him elected Professor of Natural Philosophy at Edinburgh in preference to his mentor, Sir David Brewster. His best-known scientific work was the demonstration of the polarisation of radiant heat (1834-44), and his extensive observational and theoretical work on the flow of glaciers in the 1840s. In both areas he became embroiled in controversy. In 1859, following eight years of precarious health, he moved from Edinburgh to the sinecure job of Principal of the United College at St Andrews. He died of consumption in 1868.[19] *Dissertation Sixth* was the sequel to previous *Encyclopaedia Britannica* dissertations by Forbes' predecessors in the Edinburgh chair, John Playfair and Sir John Leslie.

This paper first explores the expectations Forbes might have had of the *Encyclopaedia Britannica* as a site for discipline formation, before examining some of the broad features of his views: that physical science was led by "great men", demarcated by their moral character; his possible view that the progress of physics was periodic; and the triumvirate of mathematics-physical science-mechanical arts. In doing so, it draws also on a much briefer article that Forbes wrote two years later, "The History of Science and Some of its Lessons".[20] Unless otherwise stated, quotations are from the *Dissertation.*

## Constituting physics through the *Dissertation Sixth*

Forbes' encyclopaedia entry, *Dissertation Sixth,* provides a lens to examine physics as a cognitive, practical, and social enterprise. Like the textbooks discussed by Simon, encyclopaedias provided a nexus for the role of publishers, the book trade, readers, and men of science, in the formation of physics.[21] This section examines the opportunities the *Encyclopaedia Britannica* offered for making and managing physical science and Forbes' position within it. He exploited the ever-increasing reputation of the *Britannica* and its rapidly expanding readership, to promote his own reputation, and enlist other men of science into what was, effectively, a work of collective authorship, that portrayed science as a moral enterprise.

By 1850 the *Encyclopaedia Britannica* was becoming pre-eminent among competitors such as Coleridge's *Encyclopaedia Metropolitana*, or Brewster's *Edinburgh Encyclopaedia*.[22] All gave prominence to science and technology, and were sold in parts or by subscription, but differed in intended audience, price, and organisational structure. Lardner's *Cabinet Cyclopaedia,* for example, was part of the drive for cheap educational publications of the 1820s and 30s, aimed at the middle and wealthier working classes. Volumes of the *Cabinet Cyclopaedia*, sold for six shillings, compared to around 35 shillings a volume for the seventh edition of the *Encyclopaedia Britannica*.[23]

Unlike textbooks and book collections such as the *Library of Useful Knowledge,* encyclopaedias wrestled with the tension between providing a unified view of the entirety of knowledge, and the fragmentation necessary for readable, focused, sections, and publication as a series of issues. Most adopted some sort of alphabetical ordering though Coleridge in the *Encyclopaedia Metropolitana* attempted to retain a logical structure.[24] Since its inception, the *Encyclopaedia Britannica* had tackled the problem through a dual system of 'treatises' covering an entire subject field, and 'articles' focusing on terms and topics within a field.[25] All appeared in the same alphabetical sequence, but were distinguished by length and typography. Into this organisational system, Edinburgh publisher Archibald Constable and his editor Macvey Napier introduced a new binding element: extended "preliminary dissertations" at the beginning of each volume of the *Supplement* (1815-24) to the fourth, fifth and sixth editions, Their inspiration was D'Alembert's "Preliminary Discourse" to the *Encyclopédie*, stressing Enlightenment ideals of progress providing a coherent historical framework for a cognate body of knowledge.[26]  The first dissertation, by Dugald Stewart on the "Progress of Metaphysical, Ethical and Political Philosophy" was followed by John Playfair's on the "Progress of Mathematical and Physical Science" up to 1750 and William Brande's on the "progress of chemical philosophy". For the seventh edition of the *Encyclopaedia,* Napier and the new owners, A&C Black, added John Leslie's sequel to Playfair's *Dissertation*, and James Mackintosh's companion to Stewart, but Brande's was dropped.

Thomas Traill, professor of medical jurisprudence at Edinburgh, succeeded Napier as editor for the eighth edition in the 1850s, but little change in disciplinary approach is discernible. Traill's preface was an abridgement and brief update of Napier's. He retained the broad boundaries established by Napier when he re-numbered the dissertations and added two more to the existing four: Richard Whately's *Dissertation Third,* on the "Rise, Progress and Corruptions of Christianity", and Forbes' *Dissertation Sixth,* which formed a sequel to Playfair's and Leslie's, bringing the account of mathematical and physical science right up to date.

Traill commissioned *Dissertation Sixth* in 1852. Forbes, then in his early 40s, was away from Edinburgh for two years being treated for consumption by John Addington Symonds in Clifton, Bristol. He initially declined on health grounds - but subsequently agreed, provided he could extend the deadline and raise his fee.[27] The dissertation was eventually issued free along with volume 12 in 1856. It was published separately in 1858.[28]

By the time Traill commissioned Forbes' *Dissertation*, the *Encyclopaedia Britannica* had acquired a national, and even international, reputation. Beginning in 1815, Constable and Napier had recruited big name specialist authors, giving the London elite a stake, and abandoned the traditional anonymity of articles; the *Britannica* as a whole became the work of a collective of experts. Constable and Napier's break with the previous tradition, of encyclopaedias written largely by their editors, was a move soon copied by other major encyclopaedias such as the *Metropolitana* and the *New Cyclopaedia*.[29] By 1854 the "Encylopädie" entry in the Leipzig-based *Kleineres Brockhaus'sches Conversations Lexikon für den Handgebrauch* claimed that no encyclopaedia of scientific importance had appeared in France since 1832, but that England had numerous extensive and expensive encyclopaedias, of which the foremost was the *Britannica*.[30] Since 1790 editions of the *Britannica* had circulated widely in North America in both agreed and pirated versions.[31]

Thus Forbes would expect a sizeable home, and perhaps overseas audience. In Britain the potential readership grew as literacy rose and publication costs came down, but also as public appetite for "improvement" literature increased.[32] In Scotland, the *Encyclopaedia Britannica* was the second most borrowed title from the Leightonian Library at Dunblane, being checked out 48 times between 1780 and 1833. Its borrowers included ministers, writers, surgeons and landed gentlemen.[33] It was also listed in 151 of the 395 post-1770 library catalogues surveyed by Towsey; it seems likely that the Forbes family's well-stocked library contained a copy.[34] The Edinburgh Mechanics' Subscription Library held both the sixth and seventh editions, and Constable's *Supplements*, giving a potential readership of 1200.[35] In Britain as a whole, by the early nineteenth century non-religious improvement publications, such as the *Britannica* accounted for half of total booksellers' stock, and circulating libraries typically contained eighty per cent "standard nonfiction".[36]

But the boundaries between the educated public who read works, and the learned authors who wrote them, were blurring. In Germany, Phillips views this erosion as defined by ever-lower barriers to authorship.[37]  But the *Encyclopaedia Britannica* demonstrates, that in Britain blurring was equally defined by readership. By 1827 Napier, arguing with the new owners, A&C Black that scientific articles should not be curtailed to save money, noted that, "Encyclopaedias have risen into consequence with an important and influential class, for whose use they were not originally designed … they are now regularly perused or consulted by men of science, and the whole body of the learned."[38] Topham points out that the object of this reading was often social, to be able to discuss science in learned but general societies and fashionable parties.[39] Such reading by scientific specialists was an expectation in 1856 when *The Athenaeum*, reviewed Forbes' *Dissertation*: "Many persons, competently informed as to some of the chapters, will gain their first knowledge about the subjects of others from the chapters themselves."[40]

Through the nineteenth century, the boundary work of differentiating men of science from the wider educated public moved towards participation in specialist societies and publications.[41] But, as the example of Forbes and the *Encyclopaedia Britannica* shows, in Britain in the 1850s it was still deemed desirable for a devotee of physical science to define himself within a broad learned public. Forbes participated both in increasingly dedicated scientific societies, such as the BAAS – itself subdivided into sections - and in more generalist but "learned" societies such as the Royal Society of Edinburgh (RSE), to which he was elected in 1831. He was General Secretary 1840-60, and proposed writers and artists, as well as men of science, for fellowship.[42] He published in specifically scientific journals, but wrote also in the *Edinburgh* and *Quarterly Review*s, adhering to a publication pattern of an older generation. In his *Dissertation* Forbes was consciously writing both for a general educated readership and for men of science, audiences that paralleled his own career development through social networking and institutional position.

Forbes would not have undertaken the task unless he believed that it would enhance his reputation with both groups – the physical science men, and the general educated. He used the risk to his reputation as a bargaining counter to increase his fee from £200 to £350 – approaching a year's "fixed salary" for a Scottish Professor - and for authority to extend beyond the contracted 100 pages.[43] Among the educated public, his reputational expectations will have been set by the success of previous editions. Playfair's and Leslie's dissertations were widely quoted. The seventh edition of the *Encyclopaedia* had sold over 5000 complete sets, plus many more bought in parts. *The Athenaeum* considered it, "the most valuable digest of human knowledge that has yet appeared in Britain."[44]

Among cultivators of physical science Forbes was acutely aware of the, "obvious delicacy of dealing with contemporary or almost contemporary reputations."[45] He mitigated the risk by canvassing widely for information and views.[46] The willingness with which these men cooperated shows that they, too, considered this work an important disciplinary enterprise. While Forbes was the sole named author of the *Dissertation*, its writing was, to some extent, a collective action. Although much of the writing was Forbes' own, he included unacknowledged extracts from his correspondents; that no one complained suggests that they not only expected this, but believed that getting their views heard was more important than personal credit. The Glasgow Professor of Natural Philosophy, William Thomson, for example, seized the opportunity to establish Henry Cavendish's credentials as the founder of the mathematical theory of electricity through an ostensibly independent and authoritative witness.[47] The account he initially sent to Forbes dwelt long on Cavendish but omitted Coulomb almost entirely, and Forbes challenged this.[48] By proof stage, he had cut down Thomson's views considerably, and supplemented them with an account of Coulomb based on Whewell's *History of the Inductive Sciences*.[49] But Thomson reiterated the importance of Cavendish, sending two long additional paragraphs, which Forbes now included.[50] Humphrey Lloyd, the Dublin Natural Philosopher, engaged in a debate over the credit due to Brewster, while Edinburgh's Professor of Technology, George Wilson, contributed substantially to finessing James Watt's reputation, whose possible priority over Cavendish on the composition of water, and indebtedness to Joseph Black over latent heat, were subjects of considerable controversy.[51] The high density of footnotes in Forbes' account of Watt demonstrates his anticipation of criticism here.[52]

However, Forbes was prepared to take his own line. This is particularly evident when he draws on lesser-known Continental sources. An example is his treatment of the Italian physician, Luigi Galvani. During his Continental tour of 1844, Forbes met the physicist Silvestro Gherardi, who had examined Galvani's manuscripts thoroughly. The two men subsequently exchanged books and instruments.[53] The resultant account in *Dissertation Sixth* is detailed, and much more favourable to Galvani than was customary in British publications, preferring him to Volta.[54] Similarly, when discussing the discovery of Neptune, Forbes sought detailed information about Leverrier's background and work from his friend, the Swiss astronomer, Émile Gautier. His treatment gave far too little credit to Cambridge astronomy for Airy's liking, but Forbes did little to amend it.[55]

Remarkably, the editor, Traill, and publishers, A&C Black had little input, contrasting with the experiences of textbook authors later in the century described by Mitchell.[56] Despite the increasing use of national and international experts, for this dissertation they remained rooted in Edinburgh, "Convinced … that the supplemental essay should, like the two first parts, proceed from one of our own professors …" (Traill's emphasis).[57] Having secured Forbes' assistance, they acceded to all his suggestions: the scope and structure of the essay, the increase in fee, and that the eventual dissertation was twice the contracted length. Since Adam Black had to bear the extra cost himself, his forbearance is remarkable. It demonstrates an awareness of the growing gulf between scientific specialists and the generally educated, but also successful enrolment of the public into the value of science by the 1850s: "In regard to its limits we are not proper judges, we are influenced by commercial considerations & convenience of size which we acknowledge to be unworthy of the subject & willingly yield to your superior views."[58] The Blacks thus subordinated the commercial to the purported moral value of science discussed in the next section.

## Men of Science – managing the audience for physics

Forbes was not only defining physical science in the *Dissertation,* he was attempting to manage its audience and negotiate its cultural position. Although less dramatic than François Arago's position in post-revolutionary France, described by Levitt, Forbes faced a similar issue: "how to come to agreement in the absence of a single external authority."[59] Forbes' early experimental work on the polarization of radiant heat, was closely associated with that of Arago and Biot; he was deeply implicated in the "optical revolution" and associated downfall of Newtonian authority. The solution he implicitly set before the public in the *Dissertation* lay not so much in scientific method as in inculcating proper "spirit" among "great men" marked by their natural genius and moral qualities. "Spirit", rather than mathematical concepts, unified Forbes' science.

The impact of the imagined public with whom Forbes was negotiating is evident in his decision to organise the work around "great men". Displaying a keen appreciation of his wider readership, their interests and ways of thinking, he explained to them that this was an attempt to make the account "lively" and "escape the formality of a history of science" (p801).[60] The readers of improving literature in Scotland, and probably England also, displayed a strong preference for history and biography.[61] Forbes exploited this preference to woo his audience. The mathematician, Augustus de Morgan, writing in, *The Athenaeum* judged this a successful move: "the biographical notices and anecdotes, which form part of the body of the accounts, will give relief and heighten interest."[62]

To Traill, Forbes revealed that he was following the model of Sir James Mackintosh's *Preliminary Dissertation on the Progress of Ethical Philosophy*, for the seventh edition of the *Encyclopedia Britannica,* which had interpolated biographical anecdotes into the discussion. Traill agreed that, "Your plan of taking the most eminent in each branch, as the spindles or axes round which you are to bind the scientific fabric, is excellent," suggesting that, like Forbes, he viewed individual character as key to disciplinary progress.[63] In attributing science so emphatically to individuals, Forbes broke with his predecessors, showing a shift towards post-Enlightenment Romanticism consequent on the breakdown of Newtonian authority.[64] Unlike Playfair and Leslie, who had focused their *Preliminary Dissertations* on the rational development of concepts, stressing mathematical analysis and inductive experiment as the twin principles driving physical science, Forbes aimed, "to select the more striking land-marks of progress in each … age, and … connect them with the *character* … of all the more eminent discoverers…."(p802 my emphasis). Attribution to individuals became a British trait in the nineteenth century that acted to associate science strongly with moral worth.[65]

Differentiation from Whewell was another reason for Forbes' biographical approach.[66] Forbes saw Whewell's *History of the Inductive Sciences*, as both his main inspiration and his chief rival, judging by the remarks scattered through his correspondence and the *Dissertation*. Whewell had privileged "discoverers" in his work, but was primarily concerned with discovery status based on inductive method, rather than with character. According to Schaffer, the historiographic role assigned to discoverers by Whewell, marks a disciplinary shift from natural philosophy's belief in the power of pure induction, to physics' recognition that inductive progress depended on intuitive genius.[67] As shown in the epigraph, Forbes, writing nearly twenty years later, appears to use "physics" or "natural philosophy" merely for stylistic convenience.[68] However, any man of science worthy of inclusion in the *Dissertation* ranked as a "philosopher".[69] Whewellian themes of genius, discovery, and selection played important roles, but were trumped by character.

The choice of subjects in *Dissertation Sixth* overlaped considerably with that of Forbes' lectures in natural philosophy at Edinburgh.[70]  But the priority and ordering differs: the *Dissertation* devotes far more space to engineering and astronomy. Forbes divided the mathematical and physical sciences into seven broad branches: Physical Astronomy and Analytical Mechanics; Astronomy; Mechanics, Civil Engineering and Acoustics; Optics; Heat including Chemical

Philosophy;[71] Electricity and Magnetism. These represented his own compilation of the divisions of Playfair's, and Leslie's *Dissertations,* but with Engineering added in. However, unlike his predecessors, who had both been Professors of Mathematics before taking the Edinburgh Chair in Natural Philosophy, Forbes had little formal mathematical training and, as discussed in the penultimate section, he left pure mathematics out. He subdivided each branch further and mapped each great man to one of the subdivisions.  These were subdivisions of knowledge, but not of socio-institutional disciplines: many of his great men worked across multiple subdivisions.

Despite his explicitly broad opening definition of the field of mathematical and physical science, Forbes' implicit definition was much narrower; it encompassed only those areas of observational or experimental science to which mathematics had been applied. The study of heat, Forbes judged, had moved from chemistry to physics at the end of the eighteenth century, as it acquired quantitative laws. Dalton's atomic and gaseous theories, being quantifiable, also had, "a strongly physical aspect" (p925).[72] However, as noted by Wilson, despite his quantitative bias, Forbes gave no hint of conceptual unity between the branches of physics and shied away from the mathematical abstractions through which William Thomson and others were already seeking unification.[73]

Thus, Forbes' account was written in a transitional period before unifying concerns had taken firm hold, and he minimised such concerns, showing himself out of line with the developing, and younger, "North British" nexus described by Smith.[74]  Despite being closely involved in publication of Thomson's papers on heat in the RSE's *Transactions* and *Proceedings* from 1849 on, in his capacity as General Secretary, Forbes did not consult Thomson on the heat section of the *Dissertation*. He gave Carnot only passing mention, and included only one paragraph on Joule and the mechanical equivalent of heat, commenting privately that he deemed Thomson's admiration of Joule, "scarcely rational".[75] He concluded that, "a larger induction is still required," but that, "there is a basis of important truth in the matter which well deserves farther enquiry" (p942). When Whewell challenged the discovery status of the mechanical equivalent of heat: "I believe it rather on Wm. Thomson's authority, than because I have satisfied myself. Are you quite satisfied?" [76] Forbes responded that he thought, "pretty much as you do. I intended to speak of it with great caution, and it seems to me that I have done so."[77] The dynamical theory of heat was prominent at the BAAS meeting the following year, and the two men agreed on their reservations about it.[78]

What does unify Forbes' account is the moral and intellectual character of the men involved – the "spirit" of their enterprise. He had fashioned his own scientific identity around such spirit: glossed as disinterested "zeal" and "ardour", it was his outstanding attribute according to the majority of the sixty-one testimonials he presented when applying for the Chair at Edinburgh in 1832/33.[79] "Spirit", Forbes claimed, was far more important than the processes of science: "it is not our business here to dwell upon mere labours of *precision*, … but to show the s*pirit* in which these labours must be undertaken" (p847, Forbes' emphasis). Thus an elitism of spirit divided the great man of science, including himself, from mere labourers.[80] His *Dissertation* exemplifies a nineteenth century concern with spirit and morals that cuts across questions about discipline and field. Although defining physical science in these terms appears to fly in the face of later institutional definitions of the discipline, Phillips has shown that local societies in Germany acted as sites of collective action through which similar ideals developed among cultivators of natural science.[81] Forbes owed his own early career to the strength of his family networking in the RSE and his central role in the nascent BAAS, and we may see his definition of the physical sciences as socially rooted in associational ideals rather than in brick-and-mortar institutions.[82]

This was an account of "philosophers" as much as of physics. Forbes aimed to impress, "upon the reader... the leading facts and features of discovery in every age, together with the intellectual characteristics of the greatest minds which contributed to it" (p801).  Like Brewster before him, he resolved the tension between the focus on individuals and the increasing community emphasis on shared norms and values, by contending that philosophers were clearly distinguished by their moral qualities.[83] The most frequent of these was perseverance. Thus Fox Talbot, the photographer, was, "a Wiltshire gentleman of great ingenuity and perseverance..." (p923); the engineer, George Stephenson, was, "a man perhaps of less genius [than Richard Trevithick} but of greater sagacity and perseverance" (p883); of astronomer Francis Baily, "No amount of contrariety and failure ... was ever known to ruffle his temper, or to make his perseverance falter" (p852). Perseverance was not confined to the British; the German Bessel, Italian Galvani, and French Regnault also evinced it. But physics was not purely a handle-turning process in which perseverance was enough. Sagacity, ingenuity, originality, and genius, were also required for scientific discovery.

That the developments he described should have the status of enduring discoveries was an underlying principle. But there was a tension, which he recognised, between the discovery-based historiography that Whewell had promoted, and Forbes' own prioritisation of intellectual spirit. The faults of discoverers who fell short of the highest moral standards could be glossed over. More problematic was the inclusion of pre-eminent philosophers who had not made any significant discovery. The former Edinburgh Professor of Natural Philosophy, John Robison, was one such. "The name of Robison may perhaps not appear to be sufficiently identified with any great discovery to merit a place" (p870), started Forbes, before expending 600 words arguing for his inclusion. Forbes made his case primarily on Robison's contributions as a critic and author – known especially for his articles on "steam" and "steam engine" in the third edition of the *Encyclopaedia Britannica* - establishing first his moral claims (generous, patient, conscientious, laborious energy), and then his intellectual, "He was also a philosopher in a high sense of the word" (p870). He was, argued Forbes, "Eminently useful in forwarding the march of science ... a few more such authors ... would be cheaply purchased by the postponement of some second-rate discoveries..." (p871). For Forbes, the right intellectual spirit, and sound judgement in selecting the signal from the increasing noise of second-rate science, trumped discovery.

Forbes' portrayal is coherent with the image of men of science conveyed by the leaders of the BAAS, of whom he was one: non-sectarian, non-political, intellectual as well as moral, whose aim was to "consolidate the role of science as the dominant mode of cognition of industrial society".[84] Elitism, enforced by selection was crucial in creating this image. This was an elitism of the intellect, rather than of class or, ostensibly, of wealth. A comparison of Forbes' accounts of Lord Rosse, creator of the largest telescope in the world at the time, and James Watt, drives this point home.[85] Both had to be shown to have earned their inclusion: "…neither rank nor wealth could absolve Lord Rosse from those toils and disappointments which attend all new and original efforts.... [He] owes his success entirely to his unwearying perseverance and mechanical skill" (p863). Conversely, Watt, "…by education and habit strictly a mechanic, he had the peculiar merit of apprehending the value of theory," and, "taught men to raise the useful arts to a new dignity … to render the labours of the workshop subservient to intellectual progress" (p865). But, as implied by his frequent use of the word "genius" it *was* an elitism of birth – of nature. "Rules are of use in the humbler and more mechanical grade of subjects, but utterly unavailing in the highest. Nature herself creates discoverers".[86] Forbes spread the term genius more widely than did Whewell, and claimed no theory-experiment hierarchy, equally likely to attribute genius to either.[87] The image he conveyed aligned with the Scottish myth of the socially open "democratic intellect", as well as that of the BAAS.[88]

The distinction he drew between science conducted by rule, and that by intellectually elite discoverers may map to that Levitt draws between two forms of sociability, one that refers to authoritarian rules to guarantee its meaning, the other predicated on rational debate between equals without centralized authority.[89] Once again, Forbes was promoting associational culture as the socio-institutional basis for discovery, assuming the pre-condition of a rational intellectual elite could be met. However, he was beginning to doubt whether it could.  In recent times, "the labourers are more numerous" and all were scrambling, "for a share in the applause which arises out of some real though perhaps not very important observation."[90]  There were few capable - as he clearly believed himself to be - to judge their claims. Thus, Forbes evinced unease that associational culture was ceasing to engender the right spirit. Fifteen years previously he had noted to Whewell his experience of the amount of "humbug" in German scientific society, and now Britain was following the same path.[91] Publication in society transactions was becoming all too easy. "Societies and academies… are now more numerous but of less certain utility. There is a danger lest they become exhibition-theatres for persons of an inferior stamp...."[92] The line Forbes drew here between the practices of true "philosophers" and the culture of display of inferior science became characteristic of the North British physicists of the next generation, discussed by Morus, and Mitchell; Forbes' own careful classroom demonstration using his world class apparatus was a component of good teaching, but public sensational display was unacceptable.[93]

But despite Forbes' efforts, as a reforming professor, to provide physical and technical education, there is no suggestion in the *Dissertation* that he saw a solution to the proliferation of inferior science in institutionalized training. His account of Watt's informal education is illuminating. Watt may, or may not, have attended the chemist, Joseph Black's, lectures at Glasgow, but through his,

"intelligent spirit," he gained, "advantages … which nineteen-twentieths of enrolled students never attain…. Whilst the laboratories of the classes of Chemistry and Natural Philosophy must have been his familiar resort, his own rooms were frequented by the most intelligent students, … where subjects of science … were diligently canvassed" (pp866-7).[94] Watt's natural personal merit gained him the associations needed to succeed in physical science, and these were more important than the rules provided by formal education.

Forbes deemed education and perseverance necessary and sufficient to produce the second-rate men whose labours were essential for producing new facts or working out the consequences of new laws. But education could not produce the first-rate geniuses who progressed and defined physical science.[95] This circumvented the danger that Forbes and Whewell foresaw, that freely disseminating the rules of induction promulgated the notion that anybody could follow them and make discoveries.[96] Although associational culture might engender a proper spirit, nature also played a necessary part in addressing the issue of lack of centralised authority. While appealing to the reading tastes of the educated public, Forbes defined physical science not by *their* labours, but by the "great men who impressed the stamp of their genius at once on their age & on the Sciences which they [authored]".[97]

## The progress of physical science

Leaving now Forbes' picture of physical science as progressed and curated by great men through an associational culture, and turning to the body of physical knowledge itself, two broad features stand out. The first is his refusal to demarcate physics from mathematics and the mechanical arts, discussed in the next section. The second is that the progress of science was, probably, periodic. He deliberately articulated and argued the first in the *Dissertation.* The second is an inference from his comments in the *Dissertation*, and his 1858 article, "History of Science and Some of its Lessons". Forbes viewed science as progressed by "cautious induction" through the insights of natural geniuses. However, the fruits of induction might be exhausted before perfect theories were reached; a new way might have to be found, perhaps through the mechanical arts.

In 1858 Forbes commended the "uncompromising lesson of cautious induction which Newton… taught."[98] His brand of induction, which allowed analogy, and hypothesis provided it could be matched by experiment, but rejected Whewell's emphasis on *a priori* knowledge, has been discussed by Olson, Wilson, and Yeo. All three place Forbes in the Scottish tradition of Playfair and Stewart and distinguish his views from those of Whewell.[99]

At the start of the *Dissertation,* Forbes surveyed briefly how the components of "cautious induction" had been put in place over the previous four centuries. A preliminary phase (1450-1550) saw the development of algebra, a pre-requisite for the application of mathematics that was to follow. The next century was characterised by the triumph of observation and experiment over dogma. The century 1650-1750 demonstrated triumphantly the successful application of mathematics to accumulated observations in the work of Newton. Finally, the period 1750-1850, with which his *Dissertation* was concerned, had seen the great expansion of experiment, enabling the methods of inductive, mathematically based, science to be applied to a far wider range of phenomena than ever before, but also supplying a necessary check on unbridled mathematical abstraction.

The basic epistemological characteristic of physical sciences, for Forbes, was that they sought efficient causal explanations. In both the *Dissertation,* and his teaching, he described an inductive ascent from carefully measured observations or "facts", to highly quantitative laws such as Kepler's that related the facts mathematically, to causal theories. The theories always had to be anchored and kept in check by the facts. Pursuing his scheme, early notes for the *Dissertation* tabulated, "the periods 1650 1750 1850 in respect of, I Great Theories, II General Laws, III Important Facts". This schematic was superficially similar to Whewell's, but it lacked the "fundamental ideas" that were central to Whewell's philosophy. Whewell had developed further Kant's notion of *a priori* categories: fundamental ideas such as space and time were supplied by the mind itself, but were latent until "unfolded" through empirical experience. Conversely, they provided the organising principles for experience. Each science had a unique fundamental idea to organise its facts; for example space was the fundamental idea of geometry, and cause that of mechanics. Whewell suggested that the first law of motion was knowable *a priori* once the idea of cause, in the special form that applied to motion, i.e. force, was unfolded. This was a sticking point that Forbes could not accept and which marked his divergence from Whewell over the possibility of *a priori* knowledge.[100]

Forbes' table showed different branches of the physical sciences at different stages of progress. By 1850, mechanics and physical astronomy had their "Great Theory" of gravity, as had optics in the undulatory theory of light. But heat had not yet achieved a "Great Theory", though it did have "General Laws" of specific and latent heat, confirmation that, as discussed in the previous section, Forbes did not recognise the Thomson's unification based on energy. Acoustics was still at the stage of "Important Facts". Conversely, electricity and magnetism had theories (plural but unspecified) in the "Great Theories" column.

Forbes used the stages as explanatory factors for the pace of progress. A move from one stage to the next signalled a burst of discoveries. Progress then slowed, as it became increasingly difficult to find anything new or meaningful, until fresh insight moved the science on to the next stage. Such insights were attained only by men endowed with natural genius. By 1850, Forbes judged, many branches of physical science were suffering inevitable decline following, "... the vast steps so recently made in Optics, in Electricity, in Magnetism, in Thermotics,[101] and in Chemical principles, [which] tended of necessity to call forth such an amount of laborious detail ... as seemed to render fresh and striking originality somewhat hopeless..."(p801). At this juncture, he might have called for greater direction of science, through university institutions, to manage progress. That he did not, shows how little he considered such institutions as the central drivers of science. Among the institutions he mentioned in passing were universities, observatories, laboratories (but only the Royal Institution, and Dr Beddoes'),[102] and scientific societies, but only observatories provided any resource beyond individual salaries.

Where Forbes looked, for the onward progress of physics, was commercial engineering: Prompted, perhaps, by the success of the BAAS in persuading marine engineers and naval architects to report their data on the strength of materials for use by men of science,[103] Forbes held that, "We are continually performing experiments on a great scale and on purely commercial principles, which no individual philosopher or merely scientific society could have ventured to attempt" (p809). University institutions played no more role in this account of the progress of physics, than they did in the education of "philosophers" discussed in the previous section.

Progress became particularly difficult once the third stage, Theories, had bedded in. This was already true of astronomy. "The more that any theory of a mathematical kind, like that of Gravitation, advances to perfection... the more intense and continuous is the labour … necessary to make any advance at all" (p824). Overall, he believed that science still had a long way to go. "We have as yet made but an insignificant advance towards that completer system of Natural Philosophy of which Newton's will form but one section" (p809). Again, there is no hint of greater unification being sought, or that the completer system would be more than an aggregation of sections such as Newton's.

At this third stage Forbes took issue with Whewell's optimistic belief in the power of his version of induction.[104] He pointed out that although Bacon had shown a way to surmount the errors of dogma that had stultified the medieval period, this did not logically guarantee indefinite progress. Inductive progress might be limited by another, not yet conceived, intellectual error.

All was not necessarily lost, though. In 1858, Forbes employed two metaphors for the progress of science, of daybreak to noon, and of spring to summer. He intertwined the two by adopting the position of an Arctic navigator. Physical science was currently approaching harvest – and the limits of induction. "The glorious sun cannot rise higher than it is at the Tropic … The burden and heat of the day falls on the labourers…."[105] He took the process no further, but it is significant that both metaphors were periodic. Induction might prove limited, leading no further than the current laborious harvest, with unimagined errors preventing further progress, but a new way forward would be found and spring or daybreak and a new burst of progress would, in the end, come again.[106] The implication was clear, and it seems that Forbes, who was well known for his enthusiasm for engineering, saw alliance with the mechanical arts as one such way forward.

## The mathematics - physical science – mechanical arts triumvirate

Although he was not new in arguing for a close relation between physical science and engineering, Forbes was relatively unusual in assigning them equal intellectual value and including engineering within the field of "physical sciences". One of the barriers to progress that he continually cautioned against was inappropriate relations between mathematics, experiment, and physical science. Lack of experiment had held science back before the sixteenth century, and a too-abstract mathematics in the eighteenth. "But as [physical] knowledge advances it extends in both directions towards speculation [i.e. mathematics] and towards practical applications, but most towards the applications" (p805).

By 1750-1850, physical science had arrived at a point where it was distinguished by drawing, "far more largely upon Experiment as a means of arriving at truth than had previously been done…. science and art have been more indissolubly united than at any previous period" (p799). Here, and in the epigraph, Forbes made two claims that he clearly expected to be contested: that the relationship with the mechanical arts was worth considering; and that the relationships were reciprocal. In some sense the early BAAS had already promulgated this argument when they established Section G (mechanical science). It was seen, too, in Airy's fascination with engineering or Whewell's claims from the 1830s on about the relationship between practical art and science. [107] But Forbes took a particular line. Where these precursors insisted on the subservience of practical to theoretical science, signified by the term "mechanical *science*" for Section G, and denied the intellectual interest of its applications, Forbes referred throughout to "mechanical *arts*" asserting a more equal status, and the interest of applications. He implicitly also asserted that the scope of physical sciences included the field of engineering – warding off the threat to his own income from student fees posed by the foundation of a new Chair of Technology at Edinburgh.[108] Thus his argument was inclusive. It contrasts with that of his former student Rankine, who, as newly appointed Regius Professor of Civil Engineering and Mechanics at Glasgow, sought an academic institutional space *between* science and practice in his addresses on "harmony of theory and practice".[109]

Forbes argued that the mechanical arts were experiment on a great scale, undertaken in a spirit of grand endeavour. Civil engineering was experiment because no physical theory yet encompassed all the real-world factors that had to be taken into account. In a passage that cleverly placed Watt and Stephenson alongside the undoubted discoverer, Galileo, he argued:

> We can all readily imagine the throb of anxiety with which Galileo pointed his glasses for the first time to the moon – with which Watt saw the cylinder of his model exhausted, and the piston descend under the action of his separate condenser – and Stephenson, the stupendous iron tube at Conway resting for the first time straight as a ramrod on its two piers – these are moments of anxiety and of triumph, which place the inventor of a machine and the architect of a structure on a par with the discoverer of a planet, or with the author of a theory (p808).

Viewed as experiment, the mechanical arts were at once limited, but also essential to progress. Since, "It is not given to man to endue matter with new properties" (p808-9), technology forestalled the otherwise inevitable slowdown of inductive ascent by placing nature under new conditions, thus generating new facts. As highlighted by Marsden in the case of Rankine, these experiments took place not in an institutional laboratory, but outside, in the commercial world.[110]

Less usual, was Forbes' claim that invention was as truly intellectual as pure physics. "The masterpieces of civil engineering… are not to be compassed without inductive skill as remarkable and as truly philosophic as any effort which the man of science exerts…" (p801). The phrase "truly philosophic", signifying valid knowledge construction, countered Whewell's sharp distinction between science and art – though having made this distinction Whewell did concern himself in the "science" of engineering, for example in his book *Mechanics of Engineering* (1841).[111] Invention, suggested Forbes, was more intellectually challenging and worthy of counting as science: "It is not to be imagined that the difficulty of the problems which occupy the speculative philosopher, or the comprehensiveness of mind required for their solution, diminishes in any degree as we descend from the regions of pure science to the walks of everyday life … In fact, the former are to be regarded as the *simpler* investigations…."(p809, Forbes' emphasis).

To clinch this argument, Forbes again employed a principle of selecting the elite. He could not afford to give the impression that all mechanics counted equal to a man of science. Novelty and innovation were essential as "… such praise is only applicable when the invention is such as to call forth the qualities which distinguish the Philosopher. It is not the mere command over the agents of nature which challenges our admiration,[112] it is the foresight, the patience, the conceptive faculty, the clear-sighted and confident anticipations of what will be the results of natural laws acting in given circumstances, these circumstances being in some essential particulars *new*" (p808, Forbes' emphasis).

The second claim that Forbes argued strongly was the reciprocal relation between mathematics, physical science, and mechanical arts. Physics and the mechanical arts were intertwined intellectually as well as experimentally: "… it is quite impossible not to admit how large a share the sciences of application have had in … compelling [men] to realize certain abstract notions far from easy of conception" (p805). Elision between "physical" and "geometrical" enabled him to reject any primacy of mathematics over physics: "… with few exceptions, theorems of the greatest value and beauty have been more frequently discovered during the attempt to solve some physical or at least geometrical problem, than in comprehensive yet indefinite attempts to generalize the relations of abstract magnitude" (p806).

For Forbes, who had little formal mathematical education, mathematics had no importance as knowledge in its own right. It had value only in relation to physics, and he was wary of abstractions, such as Poisson's potential theory upon which unifying moves in heat were based, that took mathematics too far from bodily experience. Poisson, "allowed himself to be diverted … [by] constructing a system of Physics mainly founded on the applications of analysis…. The author… shows himself as a profound analyst, but adds little to our knowledge either of principles or of important results" (p825).

Forbes' views echoed those of Brewster and had their roots in the Scottish school of common sense philosophy of Dugald Stewart and Thomas Reid: mathematics was required to have its origins in sensory experience and its conclusions checked against physical reality.[113] So, although writing a dissertation on the progress of the *mathematical* and physical sciences, Forbes avoided pure mathematics, unlike Playfair and Leslie. To his readers he explained that modern advances were too technical for a popular work. To Traill he added that, "They are also remarkably destitute of circumstances of personal or historical interest..." evidencing again his concern to woo a non-scientific audience.[114]

Forbes' account of the relations of mathematics, physics, and the mechanical arts ran counter to the direction being pushed by Whewell. He rejected Whewell's belief in *a priori* knowledge and progress towards unity through increasing abstraction. Instead, he offered a way forward through practice and engineering that could guide science importantly. Where Whewell held that the mechanical arts did not justify the doing of inductive science, which should be done for its own intellectual interest, Forbes insisted that they had - through forcing nature into new configurations and impelling men to new concepts.

The *Dissertation* gave Forbes a platform to argue views on the importance of the mechanical arts developed through his teaching experience. At Edinburgh he fought to maintain both mathematics and experimentation as part of the natural philosophy course, covering topics such as heat, electricity and magnetism, the principles of machines, the theory of steam engines. He vigorously defended his right to teach technological subjects, adding civil engineering to his course when proposals for a new professorship looked like encroaching on his preserves. A letter to his successor, Peter Guthrie Tait, gives a clue to his motivation: "I would ... recommend ... the extreme undesirableness of separating systematically the Mathematical and Experimental (or popular) departments of Nat. Philosophy."[115] The experimental and technological aspects of his course were popular, brought in student fees, and met the needs of the developing Scottish industrial middle class.[116] As noted earlier, Forbes counted many of the latter among his network as he grew older, proposing several for fellowship of the RSE – though a degree of competitiveness developed as two of his protégés Lewis Gordon and Rankine succeeded successively to the Regius Chair of Civil Engineering and Mechanics at Glasgow; the inclusion of "Mechanics" in the Chair's title trespassed on natural philosophy .[117]

By insisting that the mechanical arts should be included in the physical sciences, Forbes also, intentionally or not, countered the declinist views entertained by Herschel, Whewell, Brewster and others.[118] Earlier in his career he had put forward declinist arguments himself. But in the *Dissertation* his purpose was to enrol the public in a belief in scientific endeavour, rather than to argue support for its institutions. Although he devoted a couple of paragraphs to the decline of British mathematics in the eighteenth century and its resurrection in Cambridge and Edinburgh in the nineteenth, this was limited to mathematics. The weight of the evidence (and the number of pages) presented to the reader describing the progress of physics allied to mechanical arts negated any suggestion of a wider decline: Britain clearly led other nations in the mechanical arts. Implicitly, not only had there been no significant decline, but the on-going progress of physical science had its roots in practical applications and was independent of the activities of the Cambridge coterie.

## Conclusion

Writing in the 1850s, Forbes' portrayal of physics was strongly grounded in his own experience, localised in Scotland but partaking in British and Europe-wide communities. The *Encyclopaedia Britannica* gave him a forum for shaping the shifting socio-institutional and intellectual boundaries of physics in the mid nineteenth century while preserving his own position within them. Noticeable is Forbes' negotiation across boundaries of many different types: between physical sciences and other fields; between men of science and the educated public; between new and old philosophies and historiographies. The tensions and inconsistencies necessitated by such negotiations are everywhere apparent. They serve to locate the boundaries and demonstrate the complexity of discipline formation: discovery was becoming the accepted measure of scientific worth but could not be taken as a reliable indicator; induction could not guarantee progress but abstract hypothesis was dangerous; too much communication was as bad for progress as too little; second-rate men were needed to help with the laborious work and could be trained, but first-rate geniuses had to be born; the public needed to be enrolled in support of science but associational culture was ceasing to foster intellectual spirit effectively; university institutions were no replacement for societies but he owed his own position to one.

Despite his explicitly broad definition of the field of physical science, in practice Forbes drew the confines far more closely. De Morgan's complaint that what he actually covered was, "mathematico-physical science considered both mathematically and experimentally," suggests that Forbes reflected the extent of the field of physics in a way with which educated British readers had not yet caught up.[119] However, the men of science who actively engaged with him in production of the *Dissertation* did not demur other than in small details; they seemed content with his portrayal – a portrayal that contrasts with the division between mathematics and experimental physics that had grown up in France following the decline of Laplacian physics.[120]

But setting boundaries on physics, physical science, or physical institutions, was not Forbes' primary aim. Far more important was the attempt to engage with the public and win support by extolling physics' moral value – an attempt that was common across many scientific disciplines. Interestingly, he made no equivalent effort to gain extra support for pure physics as a guide and source for technology. Technology was included as a part of (moral) physical science, but physics gained prestige by being associated with it, not by leading it.

The eighth edition of the *Encyclopaedia Britannica* has been little studied, overshadowed as it soon became by the ninth, "scholars" edition, for which Maxwell took on the role of physical sciences editor. Like Forbes, this next generation of North British physicists, based mainly in university institutions, cultivated a broad learned public, using popular periodicals and books to pursue, particularly, their non-materialist agenda.[121] The growing demand for technical education added teachers as an audience, whom they engaged through textbooks. But their attitudes to display, and the multiplication of societies that gave a voice to inferior men, were reminiscent of Forbes.[122] His *Dissertation Sixth* evidences modes of engagement and social cohesion between scientific men and the public, prevalent in the formative years of the North British group.

Forbes, portrayed the field of physics as advanced by the elitist intellectual activity of great men, endowed with natural genius and demarcated by their high moral character, necessarily supported by the labours of an increasing multitude of un-named second and third-rate men. Physics as a discipline was curated by great men, within an associational culture that engendered true intellectual spirit. But although this societal mechanism was ceasing to provide effective curation, Forbes did not see university institutions as the way forward – possibly influenced by the lessons he drew from the German universities that produced too much "humbug", and Cambridge where a critical mass of mathematicians were driving an undesirable abstraction. Instead, he implied that a closer union with the mechanical arts might provide a way forward. The irony is that ultimately it did – though not in the way Forbes argued here - as industrial needs drove the development of university physics departments in Britain.

**Acknowledgements**
I am grateful to all the participants in the "Phases of Physics" workshop, 2016, for stimulating discussions and feedback, and especially to Daniel Mitchell for on-going encouragement and suggestions. This paper has benefitted greatly from the interest and insight of two anonymous referees.

___________________________

## Notes

Abbreviations:

StA-FP  University of St Andrews, Special Collections, Forbes Papers

TCC-WP  Trinity College, Cambridge, Whewell Papers

[1] James D. Forbes, *Dissertation Sixth: Exhibiting a General View of the Progress of Mathematical and Physical Science, Principally from 1775 to 1850* (Edinburgh: A&C Black, 1856), p.799.

[2] Jed Z. Buchwald and Sungook Hong, "Physics," in David Cahan (ed.) *From Natural Philosophy to the Sciences: Writing the History of Nineteenth-Century Science* (Chicago: Chicago University Press, 2003), pp.163–95; Iwan Rhys Morus, *When Physics Became King* (Chicago: Chicago University Press, 2005), pp.22–53.

[3] Graeme Gooday, "Precision Measurement and the Genesis of Physics Teaching Laboratories," *British Journal for the History of Science*, 23 (1990): 25–51. Daniel Mitchell (Introduction to this volume) discusses both the alternative historiographies that frame my study of Forbes.

[4] P. M. Harman, "Edinburgh Philosophy and Cambridge Physics: The Natural Philosophy of James Clerk Maxwell," in P.M. Harman (ed.) *Wranglers and Physicists: Studies on Cambridge Physicists in the Nineteenth Century* (Manchester: Manchester University Press, 1985), pp.202–24; David B. Wilson, "The Educational Matrix: Physics Education at Early-Victorian Cambridge, Edinburgh and Glasgow Universities," in Harman, *Wranglers,* pp.12–48; Richard S. Olson, *Scottish Philosophy and British Physics, 1740-1870* (Princeton: Princeton University Press, 2015); Crosbie Smith and M. Norton Wise, *Energy & Empire*, (Cambridge: Cambridge University Press, 1989), pp.36-7.

[5] The scientific importance of Edinburgh in the eighteenth, century, is of course well known. Scotland's role in the resurgence of British mathematics in the nineteenth century has recently been stressed by, for example, Alex Craik, but such revisionist accounts have not yet extended to the defining of physics as a discipline. See, e.g. Alex D.D Craik, "Mathematical Analysis and Physical Astronomy in Great Britain and Ireland, 1790-1831: Some New Light on the French Connection," *Revue d'Histoire Des Mathématiques*, 22 (2016): 223–94.

[6] Jack Morrell and Arnold Thackray, *Gentlemen of Science: Early Years of the British Association for the Advancement of Science* (Oxford: Oxford University Press, 1981).

[7] Richard Yeo, "Reading Encyclopaedias: Science and the organization of knowledge in British dictionaries of arts and sciences, 1730-1850," *Isis,* 82 (1991): 24-49, 43.

[8] R. Stichweh,  *Zur Entstehung des modernen Systems wissenschaftlicher Disziplinen - Physik in Deutschland 1740–1890* (Frankfurt am Main: Suhrkamp, 1984).

[9] Steven Shapin, "The Audience for Science in Eighteenth Century Edinburgh," *History of Science* 12 (1974): 95–121. Natural history saw discipline formation through associational culture in both Germany and Scotland, see Denise Phillips, *Acolytes of Nature: Defining Natural Science in Germany, 1770-1850* (Chicago: University of Chicago Press, 2012); Diarmid A Finnegan, *Natural History Societies and Civic Culture in Victorian Scotland* (London: Pickering & Chatto, 2009).

[10] Smith and Wise, *Energy and Empire,* p.83 ; Ben Marsden, "Engineering Science in Glasgow: Economy, Efficiency and Measurement as Prime Movers in the Differentiation of an Academic Discipline," *British Journal for the History of Science*, 25 (1992): 319–46.

[11] Later in the century school teachers were a fourth important group of specialists, but not in Scotland by 1850.

[12] Thomas Neville Bonner, *Becoming a Physician: Medical Education in Britain, France, Germany, and the United States, 1750-1945* (Oxford: Oxford University Press, 1995), pp.173-4; Robert G W Anderson, "Chemistry Beyond the Academy: Diversity in Scotland in the Early Nineteenth Century," *Ambix*, 57 (2010): 84–103, 87-8; Lisa Rosner, *Medical Education in the Age of Improvement: Edinburgh Students and Apprentices, 1760-1826* (Edinburgh: Edinburgh University Press, 1991), pp.57-8, 63, 96.

[13] Josep Simon, *Communicating Physics: The Production, Circulation, and Appropriation of Ganot's Textbooks in France and England, 1851-1887* (London: Pickering & Chatto, 2011), p.213.

[14] Simon, *Communicating Physics*, pp.40-55; R. D. Anderson, *Education and Opportunity in Victorian Scotland* (Oxford: Oxford University Press, 1983), pp.70-7.

[15] James William Hudson, *The History of Adult Education: In Which Is Comprised a Full and Complete History of the Mechanics' and Literary Institutions* (London: Longman, Brown, Green & Longmans, 1851), p.vi.

[16] Hudson, *Adult Education,* pp.77-8.

[17] "Some Account of the School of Arts of Edinburgh", *Edinburgh Philosophical Journal,* 11 (1824): 203-5, 203.

[18] Lenoir has viewed discipline formation as an active remaking, by men of science, of their own culture as part of a broader reshaping of middle class culture to include scientific values. T. Lenoir, *Instituting Science: The Cultural Production of Scientific Disciplines* (Stanford: Stanford University Press, 1997), p.8.

[19] The chief source for Forbes' life is: John Campbell Shairp, Peter Guthrie Tait, and Anthony Adams-Reilly, *Life and Letters of James David Forbes* (London: Macmillan, 1873). The recent biography by Cunningham focuses on Forbes' contributions to glaciology: Frank Cunningham, *James David Forbes: Pioneer Scottish Glaciologist* (Edinburgh: Scottish Academic Press, 1990). See also: R. N. Smart, "Forbes, James David," in *Oxford Dictionary of National Biography* (Oxford:

Oxford University Press, 2004); John G Burke, "Forbes, James David," in Vol. 5 of *Dictionary of Scientific Biography* (New York: Charles Scribner's Sons, 1970-80), pp.68–9. Recent articles on specific aspects of Forbes' life include: Dennis R. Dean, "J. D. Forbes and Naples," *Geological Society, London, Special Publications*, 287 (2007): 97–107; Bruce Hevly, "The Heroic Science of Glacier Motion," *Osiris*, 11 (1996): 66–86; Nanna Kaalund, "A Frosty Disagreement: John Tyndall, James David Forbes, and the Early Formation of the X-Club," *Annals of Science*, 74 (2017): 282–98; Crosbie Smith, "William Hopkins and the Shaping of Dynamical Geology: 1830–1860," *The British Journal for the History of Science*, 22 (1989): 27–52. Forbes' election to the Edinburgh chair is discussed by: J. B. Morrell, "Brewster and the Early British Association for the Advancement of Science," in A. Morrison-Low and J. R. R. Christie (eds.) *Martyr of Science: Sir David Brewster, 1781-1868* (Edinburgh: Royal Scottish Museum, 1984), pp.25–29; Steven Shapin, "Brewster and the Edinburgh Career in Science," in *Martyr of Science,* pp.17–23.

[20] J. D. Forbes, "The history of science; and some of its lessons," *Fraser"s Magazine for Town and Country,* 57 (1858): 283-94.

[21] Simon, *Communicating Physics.*

[22] Richard Yeo, *Encylopaedic Visions: Scientific Dictionaries and Enlightenment Culture* (Cambridge: Cambridge University Press, 2001), p.170.

[23] Bernard Lightman, *Victorian Popularizers of Science: Designing Nature for New Audiences* (University of Chicago Press, 2009), p.19; Herman Kogan, *The Great EB* (Chicago: Chicago University Press, 1958), p45; Seth Rudy, "Knowledge and the Systematic Reader: The Past and Present of Encyclopedic Learning," *Culture Unbound: Journal of Current Cultural Research*, 6 (2014): 505–526, 509.

[24] Yeo, "Reading Encyclopedias," p.34.

[25] Jeff Loveland, "Unifying Knowledge and Dividing Disciplines: The Development of Treatises in the Encyclopaedia Britannica," *Book History*, 9 (2006): 57–87.

[26] Yeo, "Reading Encyclopaedias," p.33.

[27] StA-FP, ms38079/3(1), Forbes to Traill, 24 February 1852; msdep7 incoming letters 1852/25, 28, 29, Forbes and Traill letters 10, 20 & 22 March 1852.

[28] James D. Forbes, *A Review of the Progress of Mathematical and Physical Science in More Recent Times, and Particularly Between the Years 1775 and 1850* (Edinburgh: A&C Black, 1858).

[29] Yeo, *Encyclopaedic Visions*, pp.250-2.

[30] Vol. 2 of *Kleineres Brockhaussches Conversations Lexikon für den Handgebrauch* (Leipzig: Brockhaus, 1854), p.366.

[31] Kogan, *Great EB,* pp.25-6

[32] Jonathan R. Topham, "Scientific Publishing and the Reading of Science in Nineteenth-Century Britain: A Historiographical Survey and Guide to Sources," *Studies in History and Philosophy of Science Part A* 31 (2000): 559–612.

[33] Mark R. M. Towsey, "Reading the Scottish Enlightenment : Libraries, Readers and Intellectual Culture in Provincial Scotland c.1750-c.1820" (Unpublished PhD thesis, University of St Andrews, 2007), pp.162-3, 165-6.

[34] Towsey, "Reading the Scottish Enlightenment," p.32.

[35] Hudson, *Adult Education,* pp.200-1; *Laws and Catalogue of the Edinburgh Mechanics' Subscription Library*, 6th ed. (Edinburgh, 1859), pp.7-9. The Mechanics' Library also held: Lardner *Museum of Science and Art (*8 vols); *Library of Useful Knowledge* (1 vol unspecified); Partington's *British Cyclopaedia* (10 vols); *Penny Cyclopaedia* (25 vols).

[36] Vivienne S. Dunstan, "Comparing Eighteenth and Nineteenth-Century Scottish Reading Habits with England," *Review of Scottish Culture* 26 (2014): 42–8, 44.

[37] Phillips, *Acolytes*, pp.16-22.

[38] quoted in Kogan, *Great EB*, p.45.

[39] Jonathan R. Topham, "A View from the Industrial Age," *Isis* 95 (2004): 431–42.

[40] Augustus de Morgan and William Hepworth Dixon, "Reviews: *The Encyclopaedia Britannica.* Vol XII, Edinburgh, Black," *The Anthenaeum* 1521, (1856): 1563-4.

[41] Phillips, *Acolytes*; Stichweh, *Entstehung des ... Disziplinen*; James A. Secord, "How Scientific Conversation Became Shop Talk," *Transactions of the Royal Historical Society* 17 (2007): 129–56.

[42] C. D. Waterston and A. Macmillan Shearer, *Former Fellows of The Royal Society of Edinburgh, 1783-2002: Biographical Index* (Edinburgh: Royal Society of Edinburgh, 2006).

[43] StA-FP, msdep 7 incoming letters 1852/25 Forbes to Traill, 10 March 1852; msdep7 Letterbook V, pp.154-157, Forbes to Adam Black, February 1855.

[44] quoted in Kogan, *Great EB*, p.47-8.

[45] Forbes to Traill, 10 March 1852.

[46] Forbes consulted at least: his Cambridge-focused friends, John Couch Adams, George Biddell Airy, Augustus de Morgan, George Peacock, George Gabriel Stokes, William Whewell; Scotland and Ireland-based men, David Brewster; Philip Kelland, Humphrey Lloyd, William Thomson, George Wilson; engineers, Edward Sabine, Robert Stephenson; Continental workers, Émile Gautier, Christopher Hansteen, Julius Plücker.

[47] Isobel Falconer, "No actual measurement ... was required: Maxwell and Cavendish's null method for the inverse square law of electrostatics," *Studies in History and Philosophy of Science A* 65-66 (2017): 74-86.

[48] Cambridge University Library, Kelvin papers, Add7342/F213, Thomson to Forbes, 4 Jan 1854; Add7342/F214, Forbes to Thomson, 18 Jan 1854.

[49] StA-FP, msQ113.F8E8 (ms36), manuscript of dissertation on history of science by J. D. Forbes.

[50] StA-FP, msdep 7 incoming letters 1856/49, Thomson to Forbes, 16 May 1856; Forbes, *Dissertation,* paragraphs 871-2, pp.987-8.

[51] StA-FP, msdep 7 incoming letters 1856/5, Lloyd to Forbes, Jan 11 1855 [actually 1856]; for Wilson, see especially, msdep 7 incoming letters 1855/126, Wilson to Forbes, Oct 15 1855; David Philip Miller, *Discovering Water: James*

*Watt, Henry Cavendish and the Nineteenth-Century 'Water Controversy'* (Aldershot: Routledge, 2004), pp.255-8 discusses Wilson's input to Forbes' account of James Watt.

[52] Forbes, *Dissertation,* pp.866-70

[53] Silvestro Gherardi, *Opere edite ed inedite del professore Luigi Galvani, raccolte e pubblicate per cura, dell'Accademia delle Scienze dell'Instituto di Bologna* (Bologna: Emidio dell'Olmo, 1841); StA-FP msdep7 – Journals, Box 15, no.II/25 Journal of J D Forbes, 11 April to 14 September 1844, ff.47-52; Incoming letters 1846/3(a,b), 104A, Gherardi to Forbes 1846?, 8 Dec 1846; Letterbook IV pp.124-6, Forbes to Gherardi, 10 Nov 1846.

[54] Forbes, *Dissertation*, pp.958-63; cf. William Whewell, *History of the Inductive Sciences : From the Earliest to the Present Time*, 2nd edn, 3 vols (London : John W. Parker, 1847), vol.3 pp.80-6.

[55] StA-FP, msdep 7 incoming letters 1854/42, Gautier to Forbes, 25 Feb 1854; 1855/92, Airy to Forbes, 6 July 1855.

[56] Daniel Jon Mitchell, "Absolute Measurement, Elementary Physics Pedagogy, and the Reform of Dynamics in Britain, 1863-73," (this volume).

[57] StA-FP, msdep 7 incoming letters 1852/22, Traill to Forbes, 27[?] Feb 1852.

[58] StA-FP, msdep 7 incoming letters 1855/37, A&C Black to Forbes, 1 March 1855.

[59] Theresa Levitt, *Shadow of Enlightenment: Optical and Political Transparency in France 1789-1848* (Oxford: Oxford University Press, 2009), p.12.

[60] For the relationship between authors and readers in encyclopaedia publishing see, for example: Robert Darnton, *The Business of Enlightenment: A publishing history of the Encyclopédie 1775-1800* (Cambridge, Mass.: Belknap, 1979); Ulrike Spree, "How Readers Shape the Content of an Encyclopedia: A Case Study Comparing the German Meyers Konversationslexikon (1885-1890) with Wikipedia (2002-2013)," *Culture Unbound: Journal of Current Cultural Research* 6 (2014): 569–591. For the Scottish publishing context see: William Brock, 'Brewster as a Scientific Journalist', in *Martyr of Science,* pp. 37–42; Aileen Fyfe "Conscientious Workmen or Booksellers' Hacks? The Professional Identities of Science Writers in the Mid-Nineteenth Century," *Isis*, 96 (2005): 192–223; Aileen Fyfe, *Steam-Powered Knowledge: William Chambers and the Business of Publishing, 1820-1860* (University of Chicago Press, 2012); Steven Shapin "Brewster", pp.17–23.

[61] Vivienne Dunstan, *Reading Habits in Scotland circa 1750-1820* (Unpublished PhD thesis, University of Dundee, 2010), esp. pp.90, 95.

[62] De Morgan and Dixon, "Review" p.1563.

[63] StA-FP, msdep 7 incoming letters 1852/82, 83(a,b) Forbes and Traill letters, 13, 16 November 1852.

[64] Levitt, *Shadow of Enlightenment,* p.3.

[65] Richard Yeo, *Defining Science: William Whewell, natural knowledge and public debate in early Victorian Britain* (Cambridge: Cambridge University Press, 1993), p.117.

[66] Forbes to Traill, 13 November 1852.

[67] Simon Schaffer, "Scientific Discoveries and the End of Natural Philosophy," *Social Studies of Science* 16 (1986): 387–420, esp. pp.411-3.

[68] Forbes used "physics" when qualified with an adjective, e.g. "mathematical physics", "terrestrial physics". He used "Natural Philosophy" when it was the proper title of institutional positions or treatises. Where he had free choice, he used "physics" more frequently than "natural philosophy"; he capitalized Natural Philosophy more often than Physics, but it is difficult to distinguish any further rationale for when he used one or the other.

[69] Caroline Herschel and Mary Somerville had brief mentions; neither rated a section heading, but both got index entries. Maria Agnesi's *Analytical Institutions* was mentioned in a footnote, though only her family name was given, with no indication that she was female.

[70] Listed in Wilson, "Educational Matrix," p.22.

[71] This section is the only hint that Forbes might see physics as close to chemistry – contrasting with Simon's findings for France, see Simon *Communicating Physics,* p.213.

[72] Maxwell expressed an evolved version of this view twenty years later, when "mathematics" had been succeeded by "dynamics". Physics occupied that space - between "the abstract sciences of arithmetic, algebra and geometry" and chemistry - for which "dynamics" provided a fundamental explanation. Although chemistry was expanding rapidly, dynamical explanations for some chemical phenomena were "reclaiming large tracts of good ground" for physics: James Clerk Maxwell, "Physical Sciences" in Vol. 11 *Encyclopaedia Britannica,* 9th ed. (Edinburgh: A&C Black, 1885), pp.1-3.

[73] Wilson, "Educational Matrix," p.32.

[74] Crosbie Smith, *The Science of Energy: A Cultural History of Energy Physics in Victorian Britain* (Chicago: University of Chicago Press, 1998).

[75] TCC-WP Add.Ms.a/204/116, Forbes to Whewell, 2 Nov 1856.

[76] StA-FP, msdep7 incoming letters 1856/94, Whewell to Forbes, 23 Oct 1856.

[77] Forbes to Whewell, 2 Nov 1856.

[78] TCC-WP Add.Ms.a/204/121 Forbes to Whewell, 9 Sept 1857; StA-FP msdep7 incoming letters 1857/79, Whewell to Forbes 19 Sept 1857.

[79] StA-FP 38079/35(i) Bound copy of Forbes' testimonials for the Edinburgh chair.

[80] He clearly included himself among these "great men"; his discovery of the polarisation of radiant heat was included in his *Dissertation* (pp.956-7)*.

[81] Phillips, *Acolytes,* e.g. pp.9, 135.

---

[82] Forbes was deeply embedded in the RSE. His grandfather was a founding fellow; his father, uncle, and two uncles by marriage (James Skene and Colin MacKenzie) were also fellows. Between them they served on Council, the Physical Council, as curator and librarian, and as Treasurer. Forbes' elder brothers John and Charles also became fellows, but subsequent to Forbes' election. Waterston and Shearer, *Biographical Index*.

[83] D. Brewster, "Whewell"s Philosophy of the Inductive Sciences," *Edinburgh Review* 74 (1842): 265-306, 302.

[84] Morrell and Thackray, *Gentlemen of Science,* p.32.

[85] For Rosse see, for example, S. Schaffer "The Leviathan of Parsonstown: Literary technology and scientific representation," in Lenoir (ed.), *Inscribing science*, pp.182-222.

[86] Forbes, "History of Science," p.292.

[87] See Yeo, *Defining Science,* pp.121-2 for Whewell.

[88] The reality of Davie's *Democratic Intellect* in Scottish higher education in the nineteenth century has been questioned by Robert Anderson among others. But none deny that it was a powerful myth. George E. Davie, *The Democratic Intellect: Scotland and Her Universities in the Nineteenth Century* (Edinburgh: Edinburgh University Press, 1961); Anderson, *Education and Opportunity*.

[89] Levitt, *Shadow of Enlightenment*, p.47.

[90] Forbes, "History of Science," pp.284-5, 293.

[91] e.g. StA-FP, msdep7 Letterbook II pp467-69, Forbes to Whewell, 1 November 1837; Phillips, *Acolytes,* chapter 4, discusses similar concerns in Germany.

[92] Forbes, "History of Science," p.293.

[93] Iwan Rhys Morus, , "What Happened to Scientific Sensation?", *European Romantic Review* 22 (2011): 389–403; Mitchell, (this volume).

[94] A referee has pointed out that the phrase "intelligent students" may be an oblique reference to Forbes' predecessor as Professor of Natural Philosophy, John Robison.

[95] Forbes' views seem remarkably similar to those of his pupil, Maxwell, who, despite founding the archetypal British physics institution, the Cavendish Laboratory in Cambridge, practised a pedagogy based in apprenticeship and inspiration to self-discipline; see Isobel Falconer "Building the Cavendish and time at Cambridge" in Raymond Flood, Mark McCartney and Andrew Whitaker (eds.), *James Clerk Maxwell: Perspectives on his Life and Work* (Oxford University Press, 2014), pp.67-98.

[96] Yeo, *Defining Science,* p.98; see also StA-FP, msdep7 incoming letters 1855/40, Whewell to Forbes, 10 March 1855.

[97] Forbes to Traill, 13 November 1852.

[98] Forbes, "History of Science," p.290.

[99] Olson, *Scottish Philosophy;* Wilson, "The Educational Matrix"; Richard Yeo, "An Idol of the Market-Place: Baconianism in Nineteenth Century Britain," *History of Science* 23 (1985): 251–98.

[100] StA-FP, msdep7 Scientific papers, Box 27, nos.X/28; Wilson, "Educational Matrix".

[101] A term coined by William Whewell denoting that part of science that treats of heat (Oxford English Dictionary).

[102] For an account of Dr Beddoes's Laboratory see Frank AJL James, "The Subversive Humphry Davy: Aristocracy and Establishing Chemical Research Laboratories in Late Eighteenth- and Early Nineteenth-Century England," in Lissa Roberts and Simon Werrett (eds.) *Compound Histories: Materials, Governance and Production, 1760-1840* (Leiden: Brill, 2018).

[103] Ben Marsden, "The administration of the "engineering science" of naval architecture at the British Association for the Advancement of Science, 1831–1872," *Yearbook of European Administrative History* 20 (2008): 67–94.

[104] Davie, *Democratic Intellect*, positions Forbes' argument as a move to curb the pressure for Anglicisation of Scottish universities. His conclusions about Anglicisation have been queried by Anderson, *Education and Opportunity*, pp358-61, and J. B. Morrell, "Science and Scottish University Reform: Edinburgh in 1826," *British Journal for the History of Science* 6 (1972): 39.

[105] Forbes, "History of Science," p.289.

[106] This represents a modification, but not a negation, of his earlier view, discussed by Wilson, "Educational Matrix," that science was without limit and progressive, depending only on the rate at which it was mathematized.

[107] Morrell and Thackray, *Gentlemen of Science,* pp.259-60; for Airy see Ben Marsden and Crosbie Smith, *Engineering Empires: A Cultural History of Technology in Nineteenth-Century Britain* (Basingstoke: Palgrave Macmillan, 2005), chapter 1.

[108] Ben Marsden, "Engineering Science in Glasgow"; Geoffrey N. Swinney, "George Wilson's Map of Technology: Giving Shape to the "Industrial Arts" in Mid-Nineteenth-Century Edinburgh," *Journal of Scottish Historical Studies* 36 (2016): 165–90.

[109] William John Macquorn Rankine, *Introductory Lecture on the Harmony of Theory and Practice in Mechanics: Delivered to the Class of Civil Engineering and Mechanics in the University of Glasgow on Thursday, January 3, 1856* (London and Glasgow: Griffin, 1856); Marsden, "Engineering Science in Glasgow".

[110] Marsden, "Engineering Science in Glasgow," p.334-338

[111] Yeo, *Defining Science,* p.154; Ben Marsden, "'The Progeny of These Two 'Fellows'": Robert Willis, William Whewell and the Sciences of Mechanism, Mechanics and Machinery in Early Victorian Britain," *British Journal for the History of Science*, 37 (2004): 401–34; Joost Mertens, "From Tubal Cain to Faraday: William Whewell as a Philosopher of Technology," *History of Science*, 38 (2000): 321–42.

---

[112] A referee has pointed out that this is a reference to the definition of Civil Engineering given by Thomas Tredgold for the charter of the Institution of Civil Engineers ("being the art of directing the great sources of power in Nature").

[113] Richard Olson, "Scottish Philosophy and Mathematics 1750-1830," *Journal of the History of Ideas* 32 (1971): 29–44; *Scottish Philosophy,* e.g. pp.20-4, 332-6.

[114] Forbes to Traill, 13 November 1852.

[115] Wilson, "Educational Matrix", p.21, p.46n.24 quotes Forbes to Tait 17 Jan 1861.

[116] Anderson, *Education and Opportunity,* pp358-61; P. Robertson, "Scottish universities and Scottish industry 1860-1914," *Scottish Economic and Social History* 4 (1984): 39-54.

[117] Ben Marsden, '"A Most Important Trespass": Lewis Gordon and the Glasgow Chair of Civil Engineering and Mechanics, 1840-55", in Jon Agar and Crosbie Smith (eds.) *Making Space for Science: Territorial Themes in the Shaping of Knowledge* (Basingstoke: Macmillan, 1998), pp. 87–117; Marsden, "Engineering Science in Glasgow".

[118] e.g. Morrell and Thackray, *Gentlemen of Science;* L.J. Snyder, *The Philosophical Breakfast Club: Four Remarkable Men who Transformed Science and Changed the World* (New York: Broadway Books, 2011).

[119] De Morgan and Dixon, "Review," p.1563.

[120] R. Fox, and G. Weisz, "The institutional basis of French science in the nineteenth century," in R. Fox, and G. Weisz (eds.), *The Organisation of Science and Technology in France 1808-1914* (Cambridge: Cambridge University Press, 1980), pp.1-28, 26.

[121] See, e.g. Graeme Gooday, "Sunspots, Weather, and the Unseen Universe: Balfour Stewart's Anti-Materialist Representations of 'Energy' in British Periodicals," in Geoffrey Cantor and Sally Shuttleworth (eds.) *Science Serialized* (Cambridge, Mass: MIT Press, 2004), pp.111-48.

[122] Mitchell (this volume).